\documentclass[pre,bibtex,email,twocolumn,showpacs,showkeys,preprintnumbers,amsmath,amssymb,nofootinbib]{revtex4-1}

\usepackage{amsmath}
\usepackage{latexsym}

\def\[{\left\lbrack}
\def\]{\right\rbrack}

\def\({\left(}
\def\){\right)}

\newcommand{\be}{\begin{equation}}
\newcommand{\ee}{\end{equation}}
\newcommand{\ea}{\end{eqnarray}}
\newcommand{\ba}{\begin{eqnarray}}

\begin{document}

\title{Fractional Dirac Bracket and Quantization for Constrained Systems}

\author{Everton M. C. Abreu$^{a,b,c}$}
\email{evertonabreu@ufrrj.br}
\author{Cresus F. L. Godinho$^{a,b}$}
\email{crgodinho@ufrrj.br}

\affiliation{$^a$Grupo de F\' isica Te\'orica, 
Departamento de F\'{\i}sica, Universidade Federal Rural do Rio de Janeiro,\\
BR 465-07, 23890-971, Serop\'edica, RJ, Brazil\\
$^b$Centro Brasileiro de Pesquisas F\' isicas (CBPF),
Rua Dr. Xavier Sigaud 150, Urca, \\
22290-180, Rio de Janeiro, Brazil \\
${}^{c}$Departamento de F\'{\i}sica, ICE, Universidade Federal de Juiz de Fora,\\
36036-330, Juiz de Fora, MG, Brazil\\
\today\\}


\begin{abstract}
\noindent So far, it is not well known how to deal with dissipative systems.  There are many paths of investigation in the literature and none of them present a systematic and general procedure to tackle the problem.  On the other hand, it is well known that the fractional formalism is a powerful alternative when treating dissipative problems.  In this paper we propose a detailed way of attacking the issue using fractional calculus to construct an extension of the Dirac brackets in order to carry out the quantization of nonconservative theories through the standard canonical way.  We believe that using the extended Dirac bracket definition it will be possible to analyze more deeply gauge theories starting with second-class systems.
\end{abstract}

\maketitle

\pagestyle{myheadings}
\markright{\it Fractional Dirac Bracket and Quantization for Constrained Systems}

\newpage

\section{Introduction}


The Dirac approach was very popular in the nineties, when an industrial production of papers concerning methods treating constrained systems were developed.  The Dirac brackets (DB) \cite{3} were an unmodified common point between all papers about the subject.  The motivation of many works were to convert second-class systems into first-class ones.  The main objective was to obtain a gauge theory (first-class system), the holy grail for the Standard Model.  Although not so popular as before, the analysis of constrained systems still deserves some recent attentions in the literature \cite{1a,1b}.

In few words we can say that the main feature of gauge theories is the existence of constraints which fix boundaries in the phase space of gauge invariant systems to a submanifold  \cite{2}.  Dirac covered all the main issues concerning constraint systems \cite{3}, namely, a Hamiltonian approach to gauge theories and general constrained theories and, consequently, the corresponding operator quantization procedure.  Later on, the path integral method was found to be useful for quantizing gauge theories \cite{4a,4b,4c} and so-called second-class systems \cite{5a,5b}, where the conventional Poisson bracket must be replaced by the DB in the quantization procedure.

However, in constrained systems, it is possible to solve constraint equations \cite{2}.  The formalism proposed by Dirac for classical second-class constrained systems uses the DB to deal with the evolution problem.  The procedure is to apply the DB to functions of canonical variables in the unconstrained phase space, which avoid problems concerning the restriction of systems to constraints submanifolds \cite{2}.

On the other hand, there are various problems when considering classical systems besides the ones involving the quantization of second-class systems as we just have seen above.  These problems encompass nonconservative systems.  The curiosity about them is that the great majority of actual classical systems is nonconservative and nevertheless, the most advanced formalisms of classical mechanics deals only with conservative systems \cite{6a}.

Dissipation for example, is present even at the microscopic level.  There is dissipation in every non-equilibrium or fluctuating process, including dissipative tunneling \cite{7a,7c}, electromagnetic cavity radiation \cite{8a,8b} and so on.

One way to treat adequately nonconservative systems is through fractional calculus (FC) since it can be shown that, for example, a friction  force has its form resulting from a Lagrangian containing a term proportional to the fractional derivative, which is a derivative of any non-integer 
order \cite{6a}.

Non-linear dynamics is today an important subject of study in different physical and mathematical disciplines.  However, its real success and a radically new understanding of non-linear processes occurred in the last 40 years. This understanding was inspired by the discovery and insight of a new phenomenon known as dynamical chaos. The reason for that is easy to understand, since any typical system with more than one degree of freedom possesses chaotic motion for some initial conditions. We still do not know what is the measure of chaotic trajectories, but it seems that it is non-zero, and that makes the study of chaos important for constructing models of dynamical processes in nature \cite{GZ}.  

FC is one of the generalizations of classical calculus.  It has been used in several fields of science. FC provides a redefinition of the mathematical tools and it seems very useful to deal with anomalous and frictional systems.   In particular we can cite the continuous time random walk scheme as a physical counterpart example, where within the fractional approach it is possible to include external ``fields" in a straightforward manner. Also the consideration of transport in the phase space spanned by both position and velocity coordinates is possible within the same approach. Moreover, the calculation of boundary value problems is analogous to the procedure for the corresponding standard equations \cite{klafter,klafter2,klafter3,9,scalas,Hilfer2}.
Other important applications can be found investigating response functions where many studies have been reported on the phenomenon of nonexponential, power-law relaxation which is typically observed in complex systems such as dielectrics, ferroelectrics, polymers and so on. The main feature of such systems is a strong (in general, random) interaction between their components in the passage to a state of equilibrium. The FC approach to describing dynamic processes in disordered or complex systems such as relaxation or dielectric behavior in polymers or photo-bleaching recovery in biologic membranes has proved to be an extraordinarily successful tool. Some authors have proposed some fractional relaxation models to filled polymer networks and investigate the dependence of the decisive
occurring parameters on the filler content \cite{stanisvasky,Metzler}.  The study of exactly solvable fractional models of linear viscoelastic behavior is another successful field of application.  In recent years both phenomenological and molecular-based theories for the study of polymers and other viscoelastic materials came up with integral or differential equations of fractional order. Some current models of viscoelasticity based on FC are usually derived from the Maxwell model replacing the first order derivative $d/dt$ by its fractional version ${d^{\alpha}/{dt^{\alpha}}}$, where $\alpha$ is not integer \cite{Glockle}.

In this work we will use the well-known FC to analyze the well-established DB.  The objective is to construct a generalized DB capable of treating a bigger number of mechanical systems than the standard DB.

Since we believe that the FC has not been explored enough in field theory research yet, we tried to construct a self-sustained paper so that the issues are distributed as follows.  In Section II we furnish a short history about FC together with its main equations and formulations.  We will follow here the Riemann-Liouville (RL) approach.  In Section III we establish the so-called fractional variational principle, the ground stone for our cherished result.  However, we have to perform a modification of this fractional principle in order to include constrained systems.  In Section IV we analyze the same question but considering the action functional with generalized coordinates embedded in a fractional context, then we reinterpret these different initial conditions to obtain a general formulation for the Dirac description for constrained systems.  In Section V, we use the free relativistic model to apply the fractional bracket.  As usual, the conclusions, perspectives and last comments are depicted in the last Section, the sixth one. 

\section{Modified Variational Principle}

Nowadays, the inter- and multidisciplinarity among areas must be ever considered, and it therefore can be quite useful to study several problems from different areas of science (besides the ones mentioned in the last section) such as  viscoelasticity and damping, glassy condensation, diffusion and wave propagation, electromagnetism, chaos and fractals, heat transfer, biology, electronics, signal processing, robotics, system identification, genetic algorithms, percolation, modeling and identification, telecommunications, chemistry, irreversibility, control systems as well as engineering, economics and finance \cite{ten,ten2}.

It is well-known too from the current literature that the fractional approach can describe more precisely a myriad of physical systems. The formalism can be incorporated in many classical and quantum systems as described in the last section.  We believe that its use can be extended up to field theory domain.

The generalization of the concept of derivative  with non-integer values goes back to the beginning of the theory of differential calculus. Nevertheless, the development of the theory of FC is due to contributions of many mathematicians such as Euler, Liouville, Riemann, and Letnikov \cite{Old,Mill,Pod,kilbas}.

Since 1931, when Bauer \cite{Bau} showed that we cannot use the variational principle to obtain a single linear dissipative equation of motion with constant coefficients, a new horizon of possibilities has been glimpsed.  Nowadays it has been observed that in physics and mathematics the methodology necessary to understand new questions has changed towards more compact notations and powerful nonlinear and qualitative methods.
Derivatives and integrals of fractional order have been used to understand many physical applications.  For instance, questions about viscoelasticity and diffusion process may have a more detailed description when this approach is used.  
In nature, the majority of systems contains an internal damping process and the traditional approach based on energy aspects cannot be used everywhere to obtain the right equations of motion. 

So, after Bauer's corollary, Bateman \cite{Bat} proposed a procedure where multiple equations were obtained through a Lagrangian.  
Riewe \cite{6a} observed that using FC it was possible to obtain a formalism which could be used to describe both conservative and nonconservative systems.  Namely, using this approach one can obtain the Lagrangian and Hamiltonian equations of motion also for nonconservative systems. Agrawal studied a fractional variational problem \cite{Ag}.  A fractal concept applied to quantum physics has been investigated \cite{Lasa}.

The solution of a fractional Dirac equation of order $\alpha\,=\,2/3$ has been introduced \cite{Ras} and recently this subject has been revisited \cite{Dre}  

\bigskip

\subsection{The fractional calculus}

We give a short introduction to FC.  We believe that it will not provide the interested reader with all the FC tools, but we want to explain at least what is a fractional derivative.

The first way to formally introduce fractional derivatives proceeds from the repeated
differentiation of an integral power
\be
\label{A1}
\frac{d^n}{dx^n}\,x^m\,=\,\frac{m!}{(m-n)!}\,x^{m-n}\,\,.
\ee
For an arbitrary power ${\mu}$, repeated differentiation gives
\be
\label{A2}
\frac{d^n}{dx^n}\,x^{\mu}\,=\,\frac{\Gamma(\mu+1)}{\Gamma(\mu -n+1)}\,x^{\mu -n}\,\,,
\ee
with gamma functions replacing the factorial. The gamma functions allow for a
generalization to an arbitrary order of differentiation ${\alpha}$,
\be
\label{A3}
\frac{d^\alpha}{dx^\alpha}\,x^{\mu}\,=\,\frac{\Gamma(\mu+1)}{\Gamma(\mu -\alpha +1)}\,x^{\mu -\alpha}\,\,.
\ee
Of course, the objective of this work is that ${\alpha}$ can be a real number.
The extension defined by the latter equation corresponds to the RL
derivative. It is sufficient for handling functions that can be expanded in Taylor series.
A second way to introduce fractional derivatives uses the fact that the $n$-th
derivative is the inverse operation to an $n$-fold repeated integration. Basic is the
integral identity
\ba
\label{A4}
\int_a^x\,\int_a^{y_1}\,\ldots\,\int_a^{y_n -1}\,dy_n \ldots dy_1 f(y_n) \nonumber \\
\,=\,\frac{1}{(n-1)!}\,\int_a^x\,dy\,f(y)\,(x-y)^{n-1}\,\,.
\ea
A generalization of the expression allows one to define a fractional integral of
arbitrary order $\alpha$ via
\be
\label{A5}
_aD^{-\alpha}_x\,f(x)\,=\,\frac{1}{\Gamma(\alpha)}\,\int_a^x\,dy\,f(y)\,(x-y)^{\alpha -1}\,\,, \:\: (x\geq a).
\ee
A fractional derivative of an arbitrary order is defined through fractional integration
and successive ordinary differentiation. 

For the time being we will not use the $_aD^{\alpha}_x$ notation to indicate a fractional derivative.  We will return to this 
in the future.

\bigskip

After these few words about the fractional formalism, we think that it is important to justify our choice of using the RL fractional derivative instead of other very popular fractional derivatives, as the Caputo time derivative, for instance. This last one could be another way to pursuit the construction of the fractional DB.  However, we understood that Caputo is more appropriated to applications in several engineering problems due to the fact that it has a better relation with Laplace transform.  

In the last years the Caputo approach has been
favored relatively to the RL one, because it is believed
that the RL case leads to initial conditions without
physical meaning. This was contradicted by
Heymans and Podlubny \cite{hp} that studied several
cases and gave physical meaning to the RL initial
conditions \cite{oc}. 

Still trying to clarify our objectives, let us affirm that Caputo's definition  considers the differentiation inside the integral in order to solve the constant derivative problem.  For us it is an inconvenient way to describe some Lagrangian systems and consequently to obtain a consistent definition for our Dirac bracket, since we will be dealing only with field derivatives.  For our main purpose, the RL approach is more convenient than Caputo one.

However, it is well-known that several definitions of fractional derivatives and integrals exist.  For instance, Gr\"{u}nwald-Letnikov, Caputo, Weyl, Feller, Erdelyi-Kober and Riesz fractional derivatives as well as fractional Liouville operators, which have been popularized when fractional integration is performed in dynamical systems \cite{Ramia,Ramib}.   There is no equally simple definition that applies both to
functions expressed as exponentials and to functions expressed
as powers. In order to obtain a definition that is
as general as possible, in order to be possible to attack other problems too, it has become conventional \cite{6a} to use
an integral representation discovered by Liouville \cite{Liouville}
and extended by Riemann \cite{Riemann}.  This is the main reason of our choice among other formulations of FC.  Since 
we consider this work as the first step in the direction of the analysis of quantum field theories, we believe that a general 
definition of the fractional derivative is the more convenient one.  We will talk more about these perspectives in the last section.

\subsection{Modified Euler-Lagrange Equations}

Let us consider a smooth Lagrangian function.  For any smooth path $q:[a,b]\rightarrow M$ satisfying boundary conditions $q(a)=q_a$ and $q(b)=q_b,$ consider an action-like Riemann Liouville fractional integral as considered in Eq. (\ref{A5}) above \cite{Ramia,Ramib}   
\begin{equation}
\label{1}
{}_aS_t^{-\alpha}[q(\tau)]\,=\,{1\over \Gamma(\alpha)} \int_{a}^{t} L(\dot q(\tau), q(\tau),\tau)(t-\tau)^{\alpha-1}d\tau ,
\end{equation}
where $\Gamma(\alpha)$ is the Euler gamma function, with $\alpha \in (0,1]$ and $\dot q\,=\,{dq\over{d\tau}}$ is the derivative with respect to the intrinsic time $\tau \in (a,t^{\prime})$  and $t \in [t_0,t^{\prime}]$ is the time for some observer in a particular reference system. 

Notice that the Lagrangian in Eq. (\ref{1}) is weighted by 
${1\over \Gamma(\alpha)}(t-\tau)^{\alpha-1}$.
The time weighting acts as a time-dependent damping factor \cite{Ras},  
and obviously when $\alpha\rightarrow 1$ we re-obtain the usual functional 
\begin{equation}
\label{1AA}
{}_aS_t^{-\alpha}[q(\tau)]\,=\, \int_{a}^{t} L(\dot q(\tau), q(\tau),\tau) d\tau .
\end{equation} 
 
Constructing the variation of the action functional, $\delta S_{\alpha}\,=\,0$, we obtain after standard calculus the Euler-Lagrange equations associated with the fractional action integral \cite{Ramia} in Eq. (\ref{1}) that
\begin{equation}
{\partial L \over \partial q_i}\,-\,{d \over d{\tau}}\({\partial L \over \partial {\dot q_i}}\)\,-\,{1-\alpha\over t-\tau}{\partial L \over{\partial \dot q_i}}\,=\,0 ,\,\,\,\,\,i=1,\cdots, n \,\,.
\end{equation}
The Euler-Lagrange equation above, for some fractional action functional, must be obeyed.
Now we will consider the invariance in phase space
\begin{equation}
\delta \,{}_aS_t^{-\alpha}\,=\,0\,\,,
\end{equation}  
because we are interested in Dirac's quantization approach.   We 
intend to explore the same idea when constrained systems will be under consideration.


We can write, from Eq. (\ref{1AA}), that the  variation is
\begin{equation}
\label{5000}
\delta {}_aS_t^{-\alpha}\,=\,{1\over \Gamma(\alpha)}\,\delta\,\int_{a}^{t} \[p \dot q\,-\,H(p,q,\tau)\] (t-\tau)^{\alpha-1} d\tau\,=\,0
\end{equation}
so that
\begin{eqnarray}
\delta {}_aS_t^{-\alpha}&=&{1\over \Gamma(\alpha)} \int_{a}^{t} [\,(\delta L)\,(t\,-\,\tau)^{\alpha-1}\,+\,L (\delta(t-\tau)^{\alpha-1})\,]\,d\tau  \nonumber \\
&=&0 \,\,,
\end{eqnarray}
where $L=p \dot q\,-\,H(p,q,\tau)$ and the rest of the calculation is standard from the variational calculus textbooks.  The modification is due to FC formalism.  However, it is direct to deal with this additional factor.  Hence, 
after performing the variation of the Lagrangian as in Eq. (3) and of the damping factor; and isolating the coefficients for $\delta \dot{q}$ and 
$\delta \dot{p}$ that will be equal to zero, we obtain a new set of perturbed equations of motion,
\begin{eqnarray}
\dot q_i&=&{\partial H \over{\partial \dot p_i}} \\
\dot p_i&=&-{\partial H \over{\partial \dot q_i}}\,+\,{1-\alpha\over t-\tau}\,p_i,
\end{eqnarray}
which can be understood as the (fractional) Hamilton-Jacobi equations when this new action functional is considered.  It is clear that when $\alpha \rightarrow 1$ our results will turn back to the usual case. 

We will see later that the expression $\frac{1-\alpha}{t-\tau}\,p_i$ will be important in our fractional DB formulation.  The order $\alpha$ will be directly related to the fractional approach.
The presence of a fractional factor ${{1-\alpha}\over{t-\tau}}$ is responsible for the generation of a time-dependent damping into the dynamics of the system, which is very useful to study models with smooth turbulence. Furthermore it is possible to establish a relationship between the fractional Rayleigh dissipation function and the Euler-Lagrange equation \cite{Ramia},    
\begin{equation}
\label{oito}
{\partial L \over \partial q_i}\,-\,{d \over d{\tau}}\({\partial L \over \partial {\dot q_i}}\)\,-\,{\partial R \over{\partial \dot q_i}}\,=\,0 ,\,\,\,\,\,i=1,\cdots,n ,
\end{equation}
where $R$ is the fractional Rayleigh dissipation function given by
\begin{equation}
R\,=\,{1-\alpha \over t-\tau}\,L\,\,.
\end{equation}
Note that in Eq. (\ref{oito}) the dissipation function is part of the extended Euler-Lagrange equation.  
The origin of the third term is non-standard and is due to fractional analysis.

\section{Modified Variational Principle on Constrained Systems}

Now our main objective is to obtain an extended analysis which allows the quantization of classical systems with turbulence flow in field theory.  We know that the quantization of a classical field theory in a natural context is not a straightforward unique process.  The replacement of classical Poisson brackets by commutators of quantum operators cannot be carried out simultaneously for all conceivable dynamical variables without paying a price, i.e., internal obstructions will occur \cite{Pam,Bergmann}. 

In general, the commutation formalism is restricted firstly to a certain class of variables, such as the canonical coordinates of the theory.  All commutators obtained will be derived from this first set.  However  the classical theory may be substantiated in terms of any set of canonically conjugated variables in such a manner that the transition from Poisson brackets to quantum commutators leads to a ``weird" quantum theory, depending on the chosen canonical coordinates system.  

This kind of problem usually occurs when the classical theory has constraints, and the right prescription for this was first formulated by Dirac \cite{Pam} and Bergmann and Goldberg \cite{Bergmann}, where they pinpointed the right bracket algebra to be used. Thus our goal now is to extend our last result to constrained systems.  The action in Eq. (\ref{5000}) can be considered in phase space,
\begin{equation}
\label{doze}
{}_aS_t^{-\alpha}\,=\,{1\over \Gamma(\alpha)}\int_{a}^{b} \[p \dot q\,-\,\tilde{H}(p,q,\phi_a)\] (t-\tau)^{\alpha-1} d\tau
\end{equation}
where $\tilde{H}$ is
\begin{equation}
\tilde{H}\,=\,H\,+\,\lambda_a \phi_a\,\,.
\end{equation}
The question involved in such systems is that when we carried out the Legendre transformation where
$L(\dot q,q, t)$ becomes $H(p,q,t)$, and defined the canonical momenta as $p_i\,=\,{\partial L\over \partial \dot q_i}$perhaps the $N$ quantities are not all independent functions of the velocities.  We cannot
eliminate the $\dot q_i$'s and obtain $M$ constraints equations $\phi_a(q,p)\,=\,0$. 
Extending our discussion, we write the variation for Eq. (\ref{doze}) as
\begin{eqnarray}
\delta {}_aS_t^{-\alpha}&=&{1\over \Gamma(\alpha)}\delta\,\,\int_{a}^{b} \[p \dot q\,-\,\tilde{H}(p,q,\phi_a)\] (t-\tau)^{\alpha-1}d\tau \nonumber \\
&=&{1\over \Gamma(\alpha)}\,\int_{a}^{b}\Biggl[\delta p \dot q - \(\dot p - p \({{1-\alpha}\over{t-\tau}}\)\delta q\)-{{\partial {H}}\over{\partial q}}\delta q\nonumber \\
&-& {{\partial {H}}\over{\partial p}}\delta p+\lambda_a {{\partial \phi_a}\over{\partial q}}\delta q - \lambda_a {{\partial \phi_a}\over{\partial p}}\delta p \Biggr](t-\tau)^{\alpha-1}d \tau  \nonumber \\
&=&0 \,\,.
\end{eqnarray}
After some algebraic manipulations, some terms can be isolated allowing us to write the  Hamilton-Jacobi equations for the fractional constrained case,
\begin{subequations}
\label{fchj}
\begin{eqnarray}
\dot q &=& {{\partial {H}}\over{\partial p}}+\lambda_a {{\partial \phi_a}\over{\partial p}},  \label{fchja} \\
\dot p &=&-{{\partial {H}}\over{\partial q}}-\lambda_a {{\partial \phi_a}\over{\partial q}}+\,{{1-\alpha}\over{t-\tau}}\,p_i \label{fchjb}
\,\,,
\end{eqnarray}
\end{subequations}

\noindent and again we have in Eq. (\ref{fchjb}) a second term representing the fractional contribution. 

\subsection{The Fractional Dirac Bracket}

Consider a dynamical variable $\Theta[p_i,q_i,t]$; and using Eqs. (\ref{fchj}) we obtain
\begin{widetext}
\begin{eqnarray}
\label{dezoito}
{d \Theta \over{dt}}&=&{\partial \Theta \over{\partial q_i}}\dot q_i\,+\,{\partial \Theta \over{\partial p_i}}\dot p_i\,+\,{{\partial \Theta}\over{\partial t}} \nonumber \\
&=&{\partial \Theta \over{\partial q_i}}\left({{\partial {H}}\over{\partial p^i}}+\lambda_a {{\partial \phi_a}\over{\partial p^i}}\right)+{\partial \Theta \over{\partial p_i}} \left[-{{\partial {H}}\over{\partial q^i}}-\lambda_a {{\partial \phi_a}\over{\partial q^i}}+p^i\,{{1-\alpha}\over{t-\tau}}\right]\,+\,{{\partial \Theta}\over{\partial t}}\\ \nonumber
&=&\left[\left({\partial \Theta \over{\partial q_i}}{{\partial {H}}\over{\partial p^i}}-{\partial \Theta \over{\partial p_i}}{{\partial {H}}\over{\partial q^i}}\right)+\lambda_a\left({\partial \Theta \over{\partial q_i}}{{\partial {\phi_a}}\over{\partial p^i}}-{\partial \Theta \over{\partial p_i}}{{\partial {\phi_a}}\over{\partial q^i}}\right)-p_i\,{\alpha-1 \over t-\tau}{\partial \Theta \over {\partial p^i}}\right]+{\partial{\Theta}\over{\partial t}}\\ 
&=&\{\Theta,H\}+\lambda_a\{\Theta,\phi_a\}-p_i\,{\alpha-1 \over t-\tau}{\partial \Theta \over {\partial p^i}}+{\partial{\Theta}\over{\partial t}}\,\,.\nonumber
\end{eqnarray}   
\end{widetext}
\newpage
The constraints are dynamical variables too.  Then, substituting some of the constraints in Eq. (\ref{dezoito}), we have that
\begin{eqnarray}
{d \phi_a \over{dt}}&=&\{\phi_a,H\}+\lambda_b\{\phi_a,\phi_b\}-p_i\,{\alpha-1 \over t-\tau}{\partial \phi_a \over {\partial p^i}}+{\partial{\phi_a}\over{\partial t}}, \nonumber \\
\end{eqnarray}
and solving for $\lambda_a$, we finally obtain a new result for the DB in a fractional context, 
\begin{eqnarray}
\{F,G\}^*&=&\{F,G\}^{\mathrm{PB}}\,-\,\{F,\phi_a\}C^{-1}_{ab}\{\phi_b,G\} \nonumber \\
&+&\{F,\phi_a\}C^{-1}_{ab}p_i\, {\alpha-1 \over t-\tau}{{\partial \phi_b}\over{\partial p_i}}\,-\,p_i \,{\alpha-1 \over t-\tau}{{\partial F}\over{\partial p_i}}. \nonumber \\
\end{eqnarray}
Our calculations show precisely this new result as a natural extension for the DB.  We must observe that, the usual DB appears inside the fractional correction and the matrix $C_{ab}=\{\phi_a,\phi_b\}$ is the constraint matrix. It is obvious that when $\alpha \rightarrow 1$ we re-obtain the usual approach.

\section{Fractional Embedding}

Our next step is to build a general way to obtain the Dirac description for constrained systems.  For this we will consider the problem under different initial conditions.  A different and more general approach to analyze any dynamical system begins by considering the action as a function of generalized coordinates \cite{6a}.
\begin{eqnarray}
S[q(\tau),Q(\tau)]&=& \int_{a}^{b} L(q_n^r(\tau),Q_{n^{\prime}}^r(\tau),\tau) d\tau \\
q_n^r&=&(_aD^{\alpha}_t)^n x_r(t)\,,\,\,\,\,Q_{n^{\prime}}^r(\tau)\,=\,(_tD^{\alpha}_b)^{n^\prime}x_r(t)\nonumber,
\end{eqnarray}
with $r\,=\,1,2,\dots R$ coordinates considered, $n\,=\,1,2,\dots M$ is the sequential order of the derivatives for the generalized coordinates $q$ and $n^{\prime}\,=\,1,2,\dots M^{\prime}$ is the same for the coordinates $Q$.
It can be shown \cite{DuMu} that the necessary condition for an extremum of $S$ is satisfied by
\begin{equation}
{\partial L \over{\partial q_0^r}}+\sum_{n=1}^{N}(_tD^{\alpha}_b)^{n}{\partial L \over{\partial q_n^r}}+\sum_{n^{\prime}=1}^{N}(_aD^{\alpha}_t)^{n^\prime} {\partial L \over{\partial Q_{n^{\prime}}^r}}\,=\,0,
\end{equation}
and the momenta have the form
\begin{eqnarray}
p^r_n&=&\sum_{k=n+1}^{N}(_tD^{\alpha}_b)^{k-n-1}{\partial L \over{\partial q_n^r}},\nonumber \\
\pi^r_{n^{\prime}}&=&\sum_{k=n^{\prime}+1}^{N}(_aD^{\alpha}_t)^{k-n^{\prime}-1}{\partial L \over{\partial Q_{n^{\prime}}^r}}.
\end{eqnarray}

It is important to observe that we could extend the approach to a phase space just considering the usual action functional depending on the generalized fractional coordinates.  

The Dirac formalism can be easily obtained here.  It is well-known that it is useful in Lagrangian constrained systems. Now we propose its extension using the FC to encompass constrained non-conservative systems.  Of course we could define our initial conditions in a different way and consequently obtain other final expressions.  
We realize that it is a very general form to deal probably non-linear systems and other kinds of phenomena.  With this objective we define our constrained Hamiltonian, 
\begin{equation}
\tilde{H}\,=\,H\,+\,\sum_k \lambda_k \Theta_k\,+\,\sum_{k^{\prime}}v_k^{\prime}X_k^{\prime},
\end{equation}
where now the constraints are in fractional form too.  We can define them by means of the RL prescription,
\begin{subequations}
\begin{eqnarray}
\Phi_k&=&{1\over{\Gamma}(k-\alpha)}\left({d\over{dt}}\right)^k \int_a^t (t-\tau)^{k-\alpha-1}\phi_k(p,q,\tau)d\tau \\ 
X_k^{\prime}&=&{1\over{\Gamma}(k^{\prime}-\alpha)}\left(-{d\over{dt}}\right)^{k^{\prime}} \int_t^b (t-\tau)^{k^{\prime}-\alpha-1}x_{k^{\prime}}(\pi,Q,\tau)d\tau . \nonumber\\ 
\end{eqnarray}
\end{subequations}
The resulting action is
\begin{eqnarray}
S\,=\,\int_t^{t^{\prime}}dt\left(\sum_{r=1}^R \sum_{n=0}^{N-1}p_n^rq_n^r+\sum_{r=1}^R \sum_{n^{\prime}=0}^{N^{\prime}-1}\pi_{n^{\prime}}Q_{n^{\prime}}-\tilde{H}\right),\nonumber \\
\end{eqnarray}
and using of the variational principle, $\delta S\,=\,0$ again, 
we can calculate the Hamilton-Jacobi equations,
\begin{eqnarray}
\label{HJRL}
_bD^{\alpha}_tp^r_n&=&{{\partial H}\over{\partial q^r_n}}+\lambda_k{{\partial \Phi_k}\over{\partial q^r_n}}\nonumber \\
_tD^{\alpha}_bQ^r_n&=&{{\partial H}\over{\partial \pi^r_{n^{\prime}}}}+v_{k^{\prime}}{{\partial X_{k^{\prime}}}\over{\partial q^r_{n^{\prime}}}}\nonumber \\
_tD^{\alpha}_a \pi^r_{n^{\prime}}&=&{{\partial H}\over{\partial Q^r_{n^{\prime}}}}+v_{k^{\prime}}{{\partial X_{k^{\prime}}}\over{\partial Q^r_{n^{\prime}}}}\nonumber \\
_aD^{\alpha}_tq^r_n&=&{{\partial H}\over{\partial p^r_n}}+\lambda_k{{\partial \Phi_k}\over{\partial p^r_n}}\,\,.
\end{eqnarray}
This form of the Hamilton-Jacobi equations is new in the literature and introduces an extension of the Poisson bracket into the RL context presented in Ref.\cite{DuMu}.  It is natural that the next step is to obtain the proper DB expression.  One way to do that is to consider some dynamical variable $F(q_n^r,p_n^r,Q_{n^{\prime}}^r,\pi_{{\prime}}^r)$ where

\begin{eqnarray}
{{dF}\over{dt}}&=&{{\partial F}\over{\partial q_n^r}}{_a}D^{\alpha}_tq^r_n\,+\,{{\partial F}\over{\partial p_n^r}}{_b}D^{\alpha}_tp^r_n\,+\,{{\partial F}\over{\partial Q^r_{n^{\prime}}}}{_t}D^{\alpha}_bQ^r_n\,+ \nonumber \\
&+&\,{{\partial F}\over{\partial {\pi}^r_{n^{\prime}}}}{_t}D^{\alpha}_a \pi^r_{n^{\prime}}\,+\,{{\partial F}\over{\partial t}}\,\,\,,
\end{eqnarray}
and after using Eq. (\ref{HJRL}) it is straightforward to build our final and main result for the DB in the RL context, namely
\begin{eqnarray}
\label{bracket}
\{A,B\}^{\star}&=&\{A,B\}\,-\,\{A,\phi_k\}C^{-1}_{kl}\{\phi_l,B\}\,- \nonumber\\ 
&-&\{A,\chi_{k^{\prime}}\}E^{-1}_{k^{\prime}l^{\prime}}\{\chi_{l^{\prime}},B\},
\end{eqnarray}
where $C$ and $E$ are constraint matrices as in the standard Dirac constraint formalism.
The consequent quantization can be described also in the standard way as 
\be
\label{quantiz}
[A,B]\,=\,i\,\hbar\,\{A,B\}^{\star}\,\,.
\ee

Now that we have constructed a proper fractional form for the DB, we believe that the conversion methods for obtaining first-class systems from second-class ones with non-linear models can be carried out.

In the next section we will apply our result obtained in Eq. (\ref{bracket}) in a well-known and simple model, the relativistic free particle.

\section{The Relativistic Free Particle}

Our objective in this section is to study one application in the light of the fractional DB introduced in Eq. (\ref{bracket}). To fix our ideas developed before, we will consider a simple example, the relativistic free particle model, to apply the fractional embedding.   

This model is well-known, and its usual Lagrangian is given by
\begin{equation}
L\,=\,-m \sqrt{{\dot x}^2}.
\end{equation}

Using the ideas of the last sections, the action under consideration is
\begin{equation}
L\,=\,-m \sqrt{\left[({_a}D^{\alpha}{_t})^n x_r\right]^2+\left[({_t}D^{\alpha}{_b})^{n^{\prime}} x_r\right]^2} .
\end{equation}

We will restrict our calculations to the case when $\alpha\,=\,1/2$ and $n\,=\,n^{\prime}\,=\,1$.  Therefore we have
\begin{equation}
L\,=\,-m \sqrt{\left[({_a}D^{1\over2}{_t})x_r\right]^2+\left[({_t}D^{1\over2}{_b}) x_r\right]^2}
\end{equation}
or in a simpler way
\begin{equation}
L\,=\,-m \sqrt{(q_1^i)^2\,+\,(Q_1^i)^2}\,\,.
\end{equation}
Of course we could consider different orders for the derivative operator, but our main intention now is to apply the method and to show its usefulnes to canonically quantize fractional systems.

Using the definition of generalized momentum we obtain the two conjugated momenta,
\begin{subequations}
\label{33}
\begin{eqnarray}
p_0^i\,&=&\,-m\,\frac{q_1^i}{\sqrt{(q_1^i)^2+(Q_1^i)^2}} \,\,,\label{33a} \\
\pi_0^i\,&=&\,-m\,\frac{Q_1^i}{\sqrt{(q_1^i)^2+(Q_1^i)^2}}\,\,.  \label{33b}
\end{eqnarray}
\end{subequations}
The primary constraint is
\begin{equation}
\phi_1\,=\,p_0^2+\pi_0^2-m^2\approx 0\,\,.
\end{equation}
As the canonical Hamiltonian is zero, to construct the extended Hamiltonian
by Dirac's prescription we can write
\begin{equation}
\tilde{H}\,=\,\lambda(p_0^2+\pi_0^2-m^2),
\end{equation}
and since it is a first-class constraint $$\dot\phi_1\,=\,\{\phi_1,\tilde{H}\}\,=\,0\,;$$
therefore we do need to fix the gauge and our choice is
\begin{equation}
q_1^0\,+\,Q_1^0-\tau\,=\,0\,.
\end{equation}
Now we have two second-class constraints 
\begin{equation}
\{\phi_1,\phi_2\}\,=\,-2(p_0^0\,+\,\pi^0_0),
\end{equation}
and the new extended Hamiltonian can be written as
\begin{equation}
\tilde{H}\,=\,\lambda_1\left[(p_0^i)^2+(\pi_0^i)^2-m^2\right]\,+\,\lambda_2\left[(q_1^i)^0\,+\,(Q_1^i)^0-\tau\right].
\end{equation}
The time evolution of these constraints give us the correct form of the Lagrange multipliers.
The extended Hamiltonian in its final form is
\begin{equation}
\tilde{H}\,=\,{1\over{2(p_0^0\,+\,\pi_0^0)}}\big[(p_0^i)^2+(\pi_0^i)^2-m^2\big].
\end{equation}
Using the DB definition from Eq. (\ref{bracket}) we can calculate finally that
\begin{subequations}
\label{freealgebra}
\begin{eqnarray}
\{q_1^i,Q_1^j\}^{\star}&=&\{\pi_0^i,\pi_0^j\}^{\star}\,=\,0\,\,, \label{freealgebraa} \\
\{q_1^i,p_0^j\}^{\star}&=&\delta^{ij}\,-\,\frac{p_0^i\delta_0^j}{p_0^0\,-\,\pi_0^0}\,\,, \label{freealgebrab} \\
\{Q_1^i,\pi_0^j\}^{\star}&=&\delta^{ij}\,-\,\frac{\pi_0^i\delta_0^j}{p_0^0\,-\,\pi_0^0}\,\,,  \label{freealgebrac}
\end{eqnarray}
\end{subequations}
and the quantization is directly obtained using the standard Eq. (\ref{quantiz}).  We can observe that the brackets obtained above have (a kind of) expected results.  In other words, the commutative result in Eq. (\ref{freealgebraa}) is standard.  The results in Eq. (\ref{freealgebrab}) and Eq. (\ref{freealgebrac}) are standard also in the first term.  The second terms in both these equations are consequences of the fractional approach.
The result obtained in Eq. (\ref{freealgebraa}) make us think about noncommutative issues.  We will talk more about this in the next section.

We have to clarify the interested reader that here we introduced the FC to investigate nonconservative physical systems.  Consequently, we are beginning to fathom other physical features inside the fractional formalism, different from the current literature.  To help us in this task, the next step would be to investigate a solid nonconservative system like radiation damping, which is not completely understood using the standard (non-fractional) formalism.  This is a target of current research by the authors and will be published elsewhere.

\section{Conclusions and perspectives}

Our main motivation is to develop an approach based on fractional variational calculus to handle non-conservative constrained systems, since FC can be used to deal with a frictional force, for example.   In other words, our effort is to construct a fractional DB.  Consequently, we showed that it is possible to think about quantization in this scenario.  In this way we proposed two kinds of fractional formulations for the DB.  

The first one is based on the RL derivative, but it is incorporated directly into the action functional.   We obtained the Hamilton-Jacobi equations that are deformed by the fractional contribution.  Consequently the DB also has the same kind of modification.

However, this first approach does not seem to be the right one.  Therefore, we changed the formalism considering a usual form for the action, but redefining the coordinates in a generalized prescription using the fractional definition according to Riewe's prescription.  The constraints were defined in the same way and the consequence was the extension of the usual Euler-Lagrange equations of motion into a fractional scenario.    

We obtained the final form for the fractional DB which has an additional term due to the FC contribution.  We showed that the standard DB can be recovered, of course.  After this result, we believe that obtaining gauge theories for non-linear systems is now an easier task.

Other and different definitions could be used with the same objective.   For example, the generalized Euler formula, Abel or Fourier integral representation, Sonin, Letnikov, Laurent, Nekrasov and Nishimoto representation can be used.  

To apply our fractional Dirac formalism, we used an example where a $D^{1\over2}$ version of the relativistic free particle was considered.  We calculated the respective DB in the fractional embedding context, and its form results very reasonable considering the conditions imposed.  It is reasonable to see that with this first prescription for the fractional DB we cannot pinpoint all the physical features of the results.  

Geometrically speaking, fractional models provide us with a memory effect about the convolution integrals and give us some differential equations with a  bigger expressive power. This allows us to consider several different physical situations such as viscoelasticity and more abstract scenarios such as mapping using tensorial fields.  Physically we can understand the derivative of order $\alpha$ of some individual velocity $v$ as the same velocity $v_\mathrm{ob}$ from the point of view of an independent observer \cite{IP}.  

As a perspective, one possible target of research would be the noncommutative fractional dynamics.  We can observe from Eq. (\ref{freealgebraa}) that this algebra is commutative.   However, we can ask whether this is the standard pattern or whether there exists any algebra induced by FC where something like Eq. (\ref{freealgebraa}) is not true, like in string theory (using ordinary calculus) in a magnetic field background \cite{witten}.  Considering this scenario, we can ask whether the application of the symplectic formalism (where we can choose the zero-mode, like in Ref.\cite{cresus}) coupled with fractional formalism, the canonical noncommutativity can be obtained.  This is our next task, together with the construction of a Moyal-Weyl product using FC to handle problems such as noncommutative quantum mechanics where the Moyal-Weyl product will have a fractional form and maybe we can obtain interesting results comparing with the fractional quantum mechanics ones in the current literature.   Or with a noncommutative algebra obtained with the DB obtained here.  One manner that we have to re-obtain the commutativity introduces a noncommutative parameter in the original non-linear system.  We can investigate how it works in a fractional scenario.

Besides the applications proposed here, we strongly believe that quantization in a fractional context is an open area and deserves more attention.  We do not know yet the whole kind of problems that can be handled using this approach.  In general, research in gravitation, condensed matter and field theory seem to be ready to be reinterpreted using the formalism of FC. 

\section{Acknowledgment}

The authors would like to thank the referees.  Their suggestions and concerns contributed substantially to transform our paper into a better one.


\begin{thebibliography} {99}
\bibitem{3}   P. A. M. Dirac, ``{\it Lectures on Quantum Mechanics}" (Beffer Graduate School of Science, Yeshiva University, New York, 1964).

\bibitem{1a}   T. G. Philbin, New J. Phys. {\bf 12}, 123008 (2010).

\bibitem{1b}   M. Maeno, Prog. Theor. Phys. Suppl. {\bf188}, 217 (2011). 

\bibitem{2}   M. I.  Krivoruchenko,  A. A. Raduta and A. Fraessler, Phys. Rev. D {\bf 73}, 025008 (2006).

\bibitem{4a}   M. Henneaux, Phys. Rept. {\bf 126}, 1 (1985). 

\bibitem{4b}   L. D. Faddeev and A. A. Slavnov, ``{\it Gauge fields, Introduction to Quantum Theory}" (Addison-Wesley, New York, 1988).

\bibitem{4c}   M. Henneaux and C. Teitelboim, ``{\it Quantization of Constrained Systems}" (Princeton University Press, Princeton, New Jersey, 1992).  


\bibitem{5a}  I. A. Batalin and E. S. Fradkin, Phys. Lett. B {\bf 180}, 157 (1986).  

\bibitem{5b}   I. A. Batalin and E. S. Fradkin, Nucl. Phys. B {\bf 279}, 514 (1987).

\bibitem{6a}   F. Riewe, Phys. Rev. E {\bf 53}, 1890 (1996); Phys. Rev. E {\bf 55}, 3581 (1997). 


\bibitem{7a}    A. O. Caldeira and A. J. Leggett, Phys. Rev. Lett. {\bf 46}, 211 (1981); Ann. Phys. {\bf 149}, 374 (1983). 


\bibitem{7c}    J. Ankerhold, H. Grabert and G. L. Ingold, Phys. Rev. E {\bf 51}, 4267 (1995).

\bibitem{8a}   I. R. Senitzky, Phys. Rev. {\bf 119}, 670 (1960).  

\bibitem{8b}    K. W. H. Stevens, Proc. Phys. Soc. London {\bf 72}, 1027 (1958).

\bibitem{GZ}    G. Zaslavsky, ``{\it Hamiltonian Chaos and Fractional Dynamics}" (Oxford University Press, New York, 2005).

\bibitem{klafter3}    R. Metzler, E. Barkai and J. Klafter, Phys. Rev. Lett. {\bf 82}, 3563 (1999). 

\bibitem{klafter}    R. Metzler and J. Klafter, Phys. Rept. {\bf 339}, 1 (2000). 

\bibitem{Hilfer2}   R. Hilfer, Chem. Phys. {\bf 284}, 399 (2002). 

\bibitem{klafter2}   R. Metzler and J. Klafter, J. Phys. A {\bf 37}, R161 (2004). 

\bibitem{scalas}    D. Fulger, E. Scalas and G. Germano,  Phys. Rev. E {\bf 77}, 021122 (2008).

\bibitem{9}  M. A. E. Herzallah and D. Baleanu, Nonlinear Dyn. {\bf 58}, 385 (2009).

\bibitem{Metzler} R. Metzler, W. Schick, H. G. Kilian and T. F. Nonnenmacher, J. Chem. Phys. {\bf 103}, 7180 (1995).

\bibitem{stanisvasky} A. Stanislavsky1, K. Weron and J. Trzmiel, Europhys. Lett. {\bf 91}, 40003 (2010).


\bibitem{Glockle}  W. G. Glockle and T. F. Nonnenmacher,   Macromolecules {\bf 24}, 6426 (1991).


\bibitem{ten}  J. A. Tenreiro Machado, M. F. Silva, R. S. Barbosa, I. S. Jesus, C. M. Reis, M. G. Marcos and A. F. 
Galhano, ``{\it Some Applications of Fractional Calculus in Engineering}", Mathematical Problems in Engineering,  639801 (2010).

\bibitem{ten2} J. A. Tenreiro Machado, V. Kiryakova and F. Mainardi, Commun. Nonlinear Sci. Numer. Simulat.  {\bf 16},  1140 (2011).

\bibitem{Old}   K. B. Oldham and J. Spanier, ``{\it The Fractional Calculus: Theory and Applications, Differentiation and Integration to Arbitrary Order}" (Academic Press, New York, 1974).

\bibitem{Mill}   K. S. Miller and B. Ross, ``{\it An Introduction to the Fractional Calculus and Fractional Differential Equations}" (Wiley-Interscience, New York, 1993).

\bibitem{Pod}   I. Podlubny, ``{\it Fractional Differential Equations. An Introduction to Fractional Derivatives, Fractional Differential Equations, Some Methods of Their Solution and Some of Their Applications}" (Academic Press, San Diego, 1999).

\bibitem{kilbas} A. Kilbas, H. Srivastava and J. Trujillo, ``{\it Theory and Applications of Fractional Differential Equations}" (Elsevier, Amsterdam, 2006).

\bibitem{Bau}   P. S. Bauer, Proc. Natl. Acad. Sci. USA {\bf 17}, 311 (1931). 

\bibitem{Bat}   H. Bateman, Phys. Rev. {\bf 38}, 815 (1931).



\bibitem{Ag}     O. P. Agrawal, J. Math. Annal. Appl. {\bf 272}, 368 (2002).

\bibitem{Lasa}    N. Laskin, Phys. Lett. A {\bf 268(3)}, (2000) 298; Chaos {\bf 10}, 780 (2000).


\bibitem{Ras}    A. Raspini, Phys. Scripta {\bf 64}, 20 (2001).

\bibitem{Dre}    D. W. Dreisigmeyer, P. M. Young, J. Phys. A {\bf 36}, 829 (2003).









\bibitem{hp}     N. Heymans and I. Podlubny, Rheol. Acta {\bf 37}, 1 (2005).

\bibitem{oc}    M. Ortigueira and F. Coito, ``{\it The Initial Conditions of Riemann-Liouville and Caputo Derivatives}" (in the 6th EUROMECH Conference ENOC, Saint Petersburg, Russia, 2008).

\bibitem{Ramia}   R. A. El-Nabulsi, D. F. M. Torres, J. Math. Phys. {\bf 49}, 053521 (2008).


\bibitem{Ramib}   R. A. El-Nabulsi, Chaos Solitons Fract. {\bf 42}, 2614 (2009). 

\bibitem{Liouville}    J. Liouville, J. Ec. Polytech. {\bf 13}, 1 (1832); {\bf 13}, 71 (1832); {\bf 13}, 163 (1832).



\bibitem{Riemann}    B. Riemann, ``{\it Gesammelte Werke}" (Teubner, Leipzig, p. 353, 1876).







\bibitem {Pam}    P. A. M. Dirac, Can. J. Math. {\bf 2}, 129 (1950).

\bibitem{Bergmann}     P. G. Bergmann and I. Goldberg, Phys. Rev. {\bf 98}, 531 (1955).

\bibitem{DuMu}    D. Baleanu and S. I. Muslih, in ``{\it Advances in Fractional Calculus, Theoretical Developments and Applications in Physics and Enginnering}," edited by J. Sabatier {\it et al.} (Springer, Netherlands, 2007),   p. 115. 

\bibitem{IP} I. Podlubny, Fract. Calc. Appl. Anal. {\bf 5},  367 (2002).

\bibitem{witten}  N. Seiberg and E. Witten, J. High Energy Phys. {\bf 9}, 32 (1999).

\bibitem{cresus}  N. Braga and C. F. L. Godinho, Phys. Rev. D {\bf 65},085030 (2002).


\end{thebibliography}
\end{document}